\documentclass[aps,prb,reprint]{revtex4-2}
\usepackage{graphicx} % Include figure files
\usepackage{amsmath} % Useful mathematical tools
\usepackage{mathtools} % For underbrace
\usepackage{gensymb} % For degree symbol
\usepackage{bm} % bold mathi
\usepackage{natbib}
\usepackage{natmove}
\usepackage[colorlinks=true, linkcolor=blue, citecolor=blue, urlcolor=blue, linktoc=page, bookmarks=false, pdfstartview={FitH}, pdfborder={0 0 0.0 [3 3]}]{hyperref} % HyperLink
\usepackage{color} % For colored text
\usepackage{dcolumn} % Align table columns on decimal point
\newcolumntype{.}{D{.}{.}{1.3}}
\newcolumntype{-}{D{.}{.}{4.0}}
\newcolumntype{:}{D{:}{:}{12.30}}
\usepackage{makecell}
\usepackage{multirow}
\usepackage{cleveref}
\usepackage{easyReview} % Review mode
\crefname{figure}{Fig.}{Figs}
\crefname{table}{Table}{Tables}
\crefname{equation}{Eq.}{Eqs.}
\crefname{section}{Sec.}{Secs.}
\renewcommand{\today}{\number\day \space \ifcase \month \or January\or February\or March\or April\or May\or June\or July\or August\or September\or October\or November\or December\fi \space \number\year} % Date
\def\m1r{\multicolumn{1}{r}}
\begin{document}
% ===== TITLE =====
\title{Altermagnetic spin splitting and symmetry-enforced partial spin degeneracy in hexagonal MnTe}
% ===== =====
% ===== AUTHORS AND AFFILIATIONS =====
\author{Suman \surname{Rooj}}
% --------------------
\author{Nirmal \surname{Ganguli}}
\email[Contact author: ]{NGanguli@iiserb.ac.in}
\affiliation{Department of Physics, \href{https://ror.org/02rb21j89}{Indian Institute of Science Education and Research Bhopal}, Bhauri, Bhopal 462066, India}
% ===== =====
\date{\today}
% ===== ABSTRACT =====
\begin{abstract}
Besides hosting several intriguing physical properties, the recently discovered time-reversal-asymmetric antiferromagnets, known as altermagnets, hold immense promise for technologies based on spintronics. Understanding the symmetry conditions leading to the spin-splitting becomes the key to further progress in the field. Hexagonal MnTe emerges as an even-parity magnet within the altermagnet family. In this work, using {\em ab initio} density functional theory (DFT) within a combination of an appropriate exchange-correlation functional and the relevant corrections, we uncover the spin-splitting features of MnTe. Our calculations reveal the spin degeneracy to be preserved in the $k_z = 0$ and $k_y = 0$ planes, while spin-splitting is observed everywhere else in the Brillouin zone, except the nodal lines identified here. To explain these findings, we provide a comprehensive symmetry analysis based on magnetic space group theory and introduce an insightful symmetry-adapted model Hamiltonian that qualitatively describes the spin-splitting behavior in different parts of the Brillouin zone. Our calculations considering spin-orbit interaction reveal no weak ferromagnetism in MnTe. Nevertheless, we discuss plausible explanations for weak ferromagnetism and anomalous Hall effect reported from experiments. Our comprehensive analysis of the magnetic space group symmetry and the DFT results leads to a thorough understanding of altermagnetism in MnTe, paving the way for possible future technology.
\end{abstract}
% ===== =====
\maketitle
% ===== INTRODUCTION =====
\section{\label{sec:intro}Introduction}
Collinear antiferromagnets with a net zero magnetization yet a broken time-reversal-symmetry-induced spin-splitting, known as altermagnets, host several intriguing features that make them promising for technology, including spin-splitter torque and magnonics with chiral magnons showing linear dispersion \cite{KarubePRL22, BaiPRL22, SmejkalPRL23}. Recent discoveries of altermagnetic features in some antiferromagnets, including hexagonal materials like MnTe, orthorhombic materials like BiFeO$_3$ and CaMnO$_3$, and tetragonal materials like MnF$_2$ made altermagnetism a highly-persued branch of research \cite{RoojAPR23, RoojPRB25, YuanPRB20, Osumi2024MnTe, Krempasky2024Nat}; spin splitting remained elusive in direct spectroscopy for the celebrated altermagnet candidate RuO$_2$ \cite{LiuPRL24}. Besides spin-splitter torque and chiral magnons, altermagnets may host other features like magnetic multipoles, anomalous Hall effect, crystal thermal Hall effect, and crystal Nernst effect \cite{BhowalPRX24, HernandezPRL21, MnTeAHE2023, Feng2022, KluczykPRB24, SmejkalNRM22, Libor2024}.

Hexagonal MnTe is one of the few examples of experimentally confirmed altermagnetic materials that exhibit almost all of the intriguing altermagnetic features \cite{RoojAPR23, HernandezPRL21, MnTeAHE2023, Feng2022, KluczykPRB24, SmejkalNRM22, Libor2024, DongARXIV24, McClartyARXIV24, LiuMagMnTe2024PRL, GonzalezAMRMnTe2024, LeeMnTe2024PRL, Osumi2024MnTe, AlaeiARXIV24, BeyARXIV24}. Recent experiments confirmed broken Kramers degeneracy and anisotropic magnetoresistance in altermagnetic MnTe \cite{GonzalezAMRMnTe2024, LiuMagMnTe2024PRL}, although the symmetry conditions leading to spin splitting in some parts of the Brillouin zone and preserving spin degeneracy in other parts are yet to be described. Inelastic neutron scattering revealed chiral magnons and confirmed {\em g}-wave magnetism in MnTe \cite{LiuMagMnTe2024PRL}. Chiral magnons hold immense promise in the context magnon spintronics \cite{ChumakNP15, LiuNC22, WangPRL24}. As opposed to ferromagnetic origin where the magnons show a quadratic dispersion and GHz frequency, chiral magnons of altermagnetic origin may have near-linear dispersion and THz frequency \cite{SmejkalPRL23, LiuNC22}, making them particularly attractive for technology. Therefore, a thorough analysis of magnetic symmetry in MnTe, unraveling the complete picture of spin-split electron bands, which also holds for other magnetic quasiparticles, i.e., chiral magnons and their time-reversed partners, becomes imperative. Additionally, spin-orbit interaction (SOI) often leads to marginally canted magnetic moments in altermagnets, resulting in a weak ferromagnetism and an anomalous Hall effect \cite{RoojPRB25, FakhredinePRB23, Milivojevic2DM24, DevarajPRM24}. MnTe is no exception, as anomalous Hall effect and weak ferromagnetism have been reported to coexist in nominally collinear altermagnet MnTe from experiments \cite{KluczykPRB24}, calling for a thorough understanding of SOI's influence on the magnetic properties.

After briefly describing the symmetry conditions leading to spin degeneracy in conventional antiferromagnets vis \`a vis spin splitting in altermagnets, in the present work, we employ first-principles density functional theory (DFT) calculations within an appropriate exchange-correlation functional and the relevant corrections for investigating the spin-splitting behavior of MnTe in its hexagonal bulk structure. We present a comprehensive symmetry analysis based on the magnetic space group to explain our DFT results for the spin-split or spin-degenerate electronic bands. Additionally, we formulate a symmetry-adapted model Hamiltonian that captures the spin-splitting behavior for the entire Brillouin zone. Incorporating SOI, we discuss the possibility of weak ferromagnetism and anomalous Hall effect in this compound. The remainder of the article is structured as follows: We highlight the basic spin-splitting conditions in relativistic and nonrelativistic regimes in \cref{sec:symmetryprinciple}. The crystal structure and calculation methodologies are presented in \cref{sec:symmetrymethod}. In \cref{sec:results}, we thoroughly discuss our results encompassing the electronic structure and symmetry analysis. Finally, we summarize our work in \cref{sec:conclusion}.
% ===== =====
\section{\label{sec:symmetryprinciple}Symmetry conditions for spin degeneracy and spin splitting}
Antiferromagnets are traditionally characterized by no net magnetization. Most often we find the bands corresponding to the opposite magnetic sublattices to exactly overlap, having the same energy dispersion and satisfying $\varepsilon(\vec{k}, \vec{\sigma}) = \varepsilon(\vec{k}, -\vec{\sigma})$; $\varepsilon$, $\vec{k}$, and $\vec{\sigma}$ denote the energy eigenvalue, crystal momentum, and spin, respectively. In a nonmagnetic compound belonging to type-II magnetic space group (MSG) \cite{BradleySYM09Book}, combined spatial inversion ($\mathcal{P}$) and time-reversal ($\mathcal{T}$) symmetry preserve the spin degeneracy for an arbitrary $\vec{k}$-vector when the structure remains invariant under spatial inversion. For a nonmagnetic system preserving the $\mathcal{T}$ symmetry, the absence of $\mathcal{P}$ symmetry leads to the lifting of spin degeneracy in the presence of spin-orbit interaction (SOI). In contrast, predicting a spin splitting or spin degeneracy in a magnetic system requires an involved symmetry analysis. Our discussion mainly focuses on antiferromagnetic compounds in type-I or type-III MSGs \cite{YuanPRM21, BradleySYM09Book}. The spatial inversion, time reversal, and their combination impact the energy dispersion in the following way:
\begin{align}
    \mathcal{P} &: \varepsilon(\vec{k}, \vec{\sigma}) \to \varepsilon(-\vec{k}, \vec{\sigma}) \text{ and} \nonumber \\
    \mathcal{T} &: \varepsilon(\vec{k}, \vec{\sigma}) \to \varepsilon(-\vec{k}, -\vec{\sigma}), \text{ therefore}, \nonumber \\
    \mathcal{PT} &: \varepsilon(\vec{k}, \vec{\sigma}) \to \varepsilon(\vec{k}, -\vec{\sigma}).
    \label{eq:PTsymmetry}
\end{align}
The $\mathcal{PT}$ symmetry thus guarantees degeneracy of the opposite-spin bands at every $k$-point in the Brillouin zone. A suitable real-space translation $\vec{\tau}$ helping in preserving the $\mathcal{PT}$ symmetry does not disturb the band degeneracy, as the real-space translation does not impact the momentum or spin space in \cref{eq:PTsymmetry}.

Considering a system without SOI, where the spatial and spin degrees of freedom are decoupled, antiferromagnetic systems will preserve spin degeneracy if a pure spin rotation operation $\mathcal{U} \in$ SU(2) leaves the system invariant:
\begin{equation}
    \mathcal{U} : \varepsilon(\vec{k},\vec{\sigma}) \to \varepsilon(\vec{k},-\vec{\sigma}). \label{eq:spinRotation}
\end{equation}
Once again, a suitable real-space translation $\vec{\tau}$ helping in preserving the $\mathcal{U}$-symmetry is acceptable, as it does not impact the momentum or spin in \cref{eq:spinRotation}.

While antiferromagnetic materials having either one of the above-mentioned spin-reversing symmetries will preserve spin-degeneracy throughout the Brillouin zone owing to every $\vec{k}$-vector remaining invariant under them, their absence does not guarantee spin-splitting everywhere in the Brillouin zone. Taking an example of MnTe, we discuss the implications of the magnetic space group symmetries in different parts of the Brillouin zone and tally them with our DFT results for a comprehensive theoretical understanding of altermagnetic spin splitting.
% ===== METHOD =====
\section{\label{sec:symmetrymethod}Crystal structure, symmetry operations, and methodology}
% ===== Figure: Structure =====
\begin{figure}
	\includegraphics[scale = 0.24]{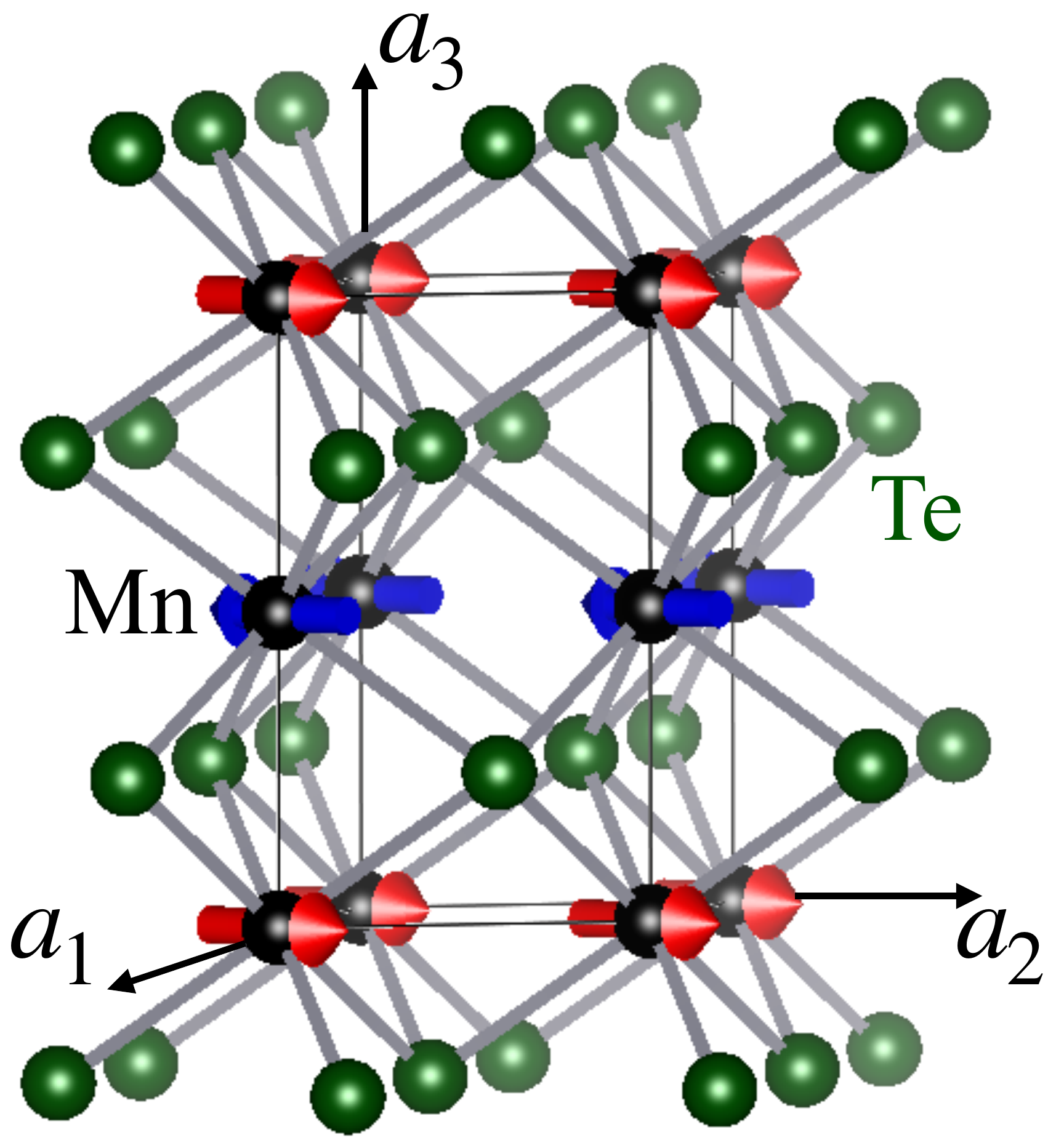}
	\caption{\label{fig:MnTestructure}The unit cell of MnTe with A-type antiferromagnetic order is illustrated in this figure.}
\end{figure}
% ===== =====
Bulk MnTe crystallizes in a hexagonal structure having the space group $P6_{3}/mmc$ with point group $D_{6h}$. \cref{fig:MnTestructure} depicts the unit cell of MnTe where Mn and Te occupy $2a$ and $2c$ Wyckoff positions, respectively.
% ===== Table: Space Group =====
\begin{table}
\caption{\label{tab:NMSpaceGroup}All symmetry operations of the space group $P6_3/mmc$ and their impact on a position vector ($x, y, z$) in the primitive lattice vector basis are listed here. We have adopted Seitz symbols for the crystallographic symmetry operations \cite{GlazerAC14, Bilbao2006, Aroyoxo5013}.}
\begin{ruledtabular}
    \begin{tabular}{:}
    \multicolumn{1}{c}{Symmetry operations} \\
    \hline
   \{1|0\} : (x, y, z) \to (x, y, z) \\
   \{3^+_{001}|0\} : (x, y, z) \to (-y, x-y, z) \\
   \{3^-_{001}|0\} : (x, y, z) \to (-x+y, -x, z) \\
   \{2_{001}|0~0~\frac{1}{2}\} : (x, y, z) \to (-x, -y, z+\frac{1}{2}) \\
   \{6^+_{001}|0~0~\frac{1}{2}\} : (x, y, z) \to (x-y, x, z+\frac{1}{2}) \\
   \{6^-_{001}|0~0~\frac{1}{2}\} : (x, y, z) \to (y, -x+y, z+\frac{1}{2}) \\
   \{2_{100}|0\} : (x, y, z) \to (x-y, -y, -z) \\
   \{2_{110}|0\} : (x, y, z) \to (y, x, -z) \\
   \{2_{010}|0\} : (x, y, z) \to (-x, -x+y, -z) \\
   \{2_{1\bar{1}0}|0~0~\frac{1}{2}\} : (x, y, z) \to (-y, -x, -z+\frac{1}{2}) \\
   \{2_{120}|0~0~\frac{1}{2}\} : (x, y, z) \to (-x+y, y, -z+\frac{1}{2}) \\
   \{2_{210}|0~0~\frac{1}{2}\} : (x, y, z) \to (x, x-y, -z+\frac{1}{2}) \\
   \{-1|0\} : (x, y, z) \to (-x, -y, -z) \\
   \{-3^+_{001}|0\} : (x, y, z) \to (y, -x+y, -z) \\
   \{-3^-_{001}|0\} : (x, y, z) \to (x-y, x, -z) \\
   \{m_{001}|0~0~\frac{1}{2}\} : (x, y, z) \to (x, y, -z+\frac{1}{2}) \\
   \{-6^+_{001}|0~0~\frac{1}{2}\} : (x, y, z) \to (-x+y, -x, -z+\frac{1}{2}) \\
   \{-6^-_{001}|0~0~\frac{1}{2}\} : (x, y, z) \to (-y, x-y, -z+\frac{1}{2}) \\
   \{m_{100}|0\} : (x, y, z) \to (-x+y, y, z) \\
   \{m_{110}|0\} : (x, y, z) \to (-y, -x, z) \\
   \{m_{010}|0\} : (x, y, z) \to (x, x-y, z) \\
   \{m_{1\bar{1}0}|0~0~\frac{1}{2}\} : (x, y, z) \to (y, x, z+\frac{1}{2}) \\
   \{m_{120}|0~0~\frac{1}{2}\} : (x, y, z) \to (x-y, -y, z+\frac{1}{2}) \\
   \{m_{210}|0~0~\frac{1}{2}\} : (x, y, z) \to (-x, -x+y, z+\frac{1}{2}) \\
    \end{tabular}
\end{ruledtabular}
\end{table}
% ===== =====
The space group comprises 24 unitary symmetry operations ($G \equiv G_U$) listed in \cref{tab:NMSpaceGroup}, adopting Seitz symbols \cite{GlazerAC14, Bilbao2006, Aroyoxo5013}. To briefly describe the notation, $1$ and $-1$ are the identity and the spatial inversion operations, respectively; $3^+_{001}$ and $3^-_{001}$ are the three-fold ($2\pi/3$) anticlockwise and clockwise rotation about the [001] axis, respectively; $2_{001}, 2_{110}, 2_{100}, 2_{010}, 2_{1\bar{1}0}, 2_{120}, 2_{210}$ are the two-fold ($\pi$) anticlockwise rotation about the [001], [110], [100], [010], $[1\bar{1}0]$, [120], [210] axes, respectively; $6^+_{001}~\text{and}~6^-_{001}$ are the six-fold ($\pi/3$) anticlockwise and clockwise rotations about [001] axis; $-3^+_{001}, -6^+_{001}, -3^-_{001}, -6^-_{001}$ represent the same operations described above, followed by inversion; $m_{001}, m_{110}, m_{100}, m_{010}, m_{1\bar{1}0}, m_{120}, m_{210}$ are the mirror operations about [001], [110], [100], [010], $[1\bar{1}0]$, [120], [210] planes, respectively. The vector $(0~0~\frac{1}{2})$ represents a non-primitive translation by $c/2$ along the [001] direction.

While $P6_3/mmc$ space group preserves inversion symmetry, all the local magnetic moments reverse their direction upon the $\mathcal{PT}$ operation. A suitable real-space translation in a conventional centrosymmetric antiferromagnet restores the local magnetic moments to the original direction, preserving $\mathcal{T}$ and $\mathcal{PT}$ symmetries. Examining the magnetic structure of MnTe depicted in \cref{fig:MnTestructure}, we find the local magnetic moments flipped by $\mathcal{T}$ or $\mathcal{PT}$ can be restored upon a translation by $\vec{a}_3/2$. However, this translation does not leave the structure invariant. In fact, a screw operation $\{2_{001}|0~0~\frac{1}{2}\}$ leaves the nonmagnetic structure invariant (see \cref{tab:NMSpaceGroup}), precluding a structural invariance upon a $\vec{a}_3/2$ translation, thus preventing $\mathcal{T}$ and $\mathcal{PT}$ symmetry to hold in the magnetic structure. The same reasoning holds for the nonrelativistic $\mathcal{U}$ operation, allowing for altermagnetic spin splitting in MnTe. Hence, the nonsymmorphic screw symmetry in the space group may be termed as a necessary condition for realizing altermagnetism in MnTe.

The electronic structure and spin-orbit interaction (SOI) calculations are performed within the framework of density functional theory (DFT) implemented in the {\scshape vasp} code \cite{vasp1, vasp2}. The projector augmented wave (PAW) method is employed for describing the potential alongside a plane wave basis set with a 500~eV energy cutoff for expanding the wavefunctions \cite{paw}. The exchange-correlation functional is approximated within the meta-generalized gradient approximation (meta-GGA) SCAN \cite{scan}, Hubbard-$U$ correction \cite{DudarevPRB98}, and van der Waals correction rVV10 \cite{rVV10} for the DFT calculations of bulk hexagonal MnTe, as optimized earlier \cite{RoojAPR23}. The integration over the Brillouin zone is performed using a $9 \times 9 \times 6$ $\Gamma$-centered $k$-point mesh for the unit cell within the corrected tetrahedron method \cite{BlochlPRB94T}. To take care of the correlation effect due to Mn-$3d$ states, we consider the Hubbard-$U$ correction with $U_\text{eff} = U - J = 3$~eV. Atomic positions and lattice vectors are optimized by minimizing the Hellman-Feynman force on each atom up to a threshold value of $10^{-2}$~eV/\AA.
%==============================================
% ===== RESULTS AND DISCUSSIONS =====
\section{\label{sec:results}Results and discussions}
% ===== Figure: Band Structure =====
\begin{figure}
	\includegraphics[scale = 0.36]{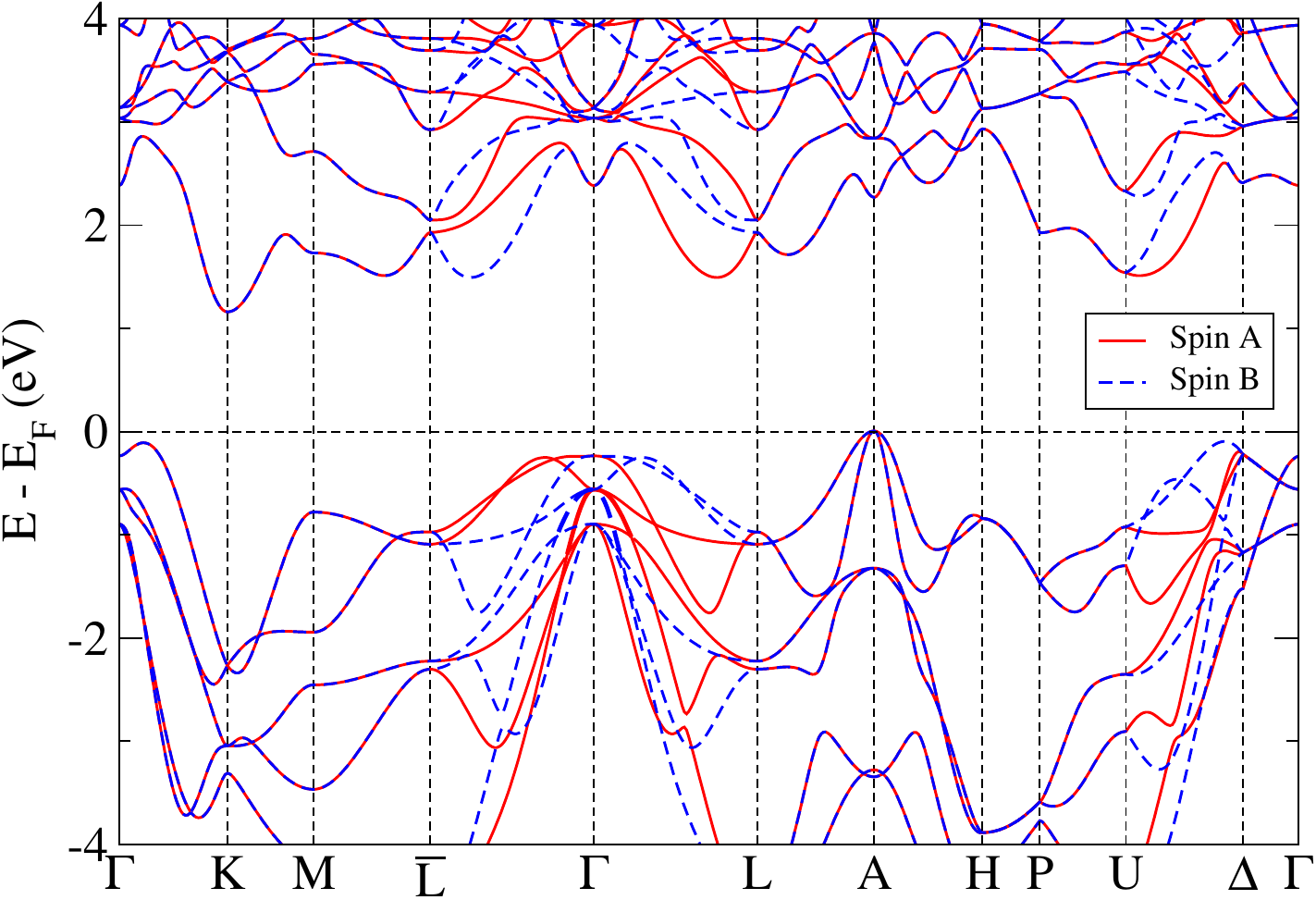}
	\caption{\label{fig:fullpath}Spin-polarized band dispersion for MnTe from both spins marked as Spin A and Spin B are depicted here. Spin degeneracy prevails except for $\bar{L} (0, 1/2, -1/2) \to \Gamma \to L (0, 1/2, 1/2)$ line and $U (0, 1/2, 1/4) \to \Delta (0, 0, 1/4)$ high symmetry lines. Reciprocal coordinates of the high-symmetry points have been mentioned in the parentheses.}
\end{figure}
% ===== =====
Below, we discuss the results obtained from DFT calculations and symmetry analysis within magnetic space group theory, first without accounting for SOI and then considering SOI.
% ===== Subsection: Electronic structure =====
\subsection{\label{subsec:Elecproperties}Electronic structure}
Hexagonal MnTe hosts antiferromagnetism with an A-type arrangement of the local magnetic moments \cite{RoojAPR23}, as illustrated in \cref{fig:MnTestructure}, with the local magnetic moments parallel to each other in the hexagonal basal plane and antiparallel along the $\vec{a}_3$-direction. \Cref{fig:fullpath} shows the electronic band structure obtained from our spin-polarized DFT calculations within the antiferromagnetic configuration of MnTe, revealing an indirect band gap of $\sim$1.16~eV between the valence band maximum at the A-point and conduction band minimum at the K-point. The band gap value reasonably agrees with 1.27~eV reported from experiment \cite{SzuszkiewiczPRB06}, in view of a systematic underestimation from DFT calculations. The projected magnetic moment of 4.54~$\mu_B$ at each Mn atom also agrees well with the experimentally measured magnetic moment of 4.7$-$5~$\mu_B$ per Mn site \cite{KriegnerPRB17}, as the projection scheme within a plane-wave basis set often leads to underestimation. The spin-polarized band dispersion for MnTe, shown in \cref{fig:fullpath}, reveals a perfect overlap of both spin bands along some of the high-symmetry directions, except $\bar{L} \to \Gamma \to L$ and $U \to \Delta$ directions. In the remainder of the article, we critically examine spin-splitting and spin-degeneracy in different parts of the Brillouin zone.
% ===== Table: Magnetic Space Group w/o SOI =====
\begin{table*}
\caption{\label{tab:MSGSpaceGroup}All symmetry operations for the magnetic space group $P6'_3/m'mc'$, which describes MnTe's magnetic symmetry when spin-orbit interaction is not considered, are listed here. ($x, y, z$) and ($u, v, w$) denote generic position and momentum vectors with respect to the primitive lattice vector basis set and the reciprocal lattice vector basis set, respectively. The transformations of ($u, v, w$) are obtained using
 \cref{eq:ConvertOperator}.}
\begin{ruledtabular}
    \begin{tabular}{::}
    %\hline
    %\hline
    \multicolumn{1}{c}{$G_U$ (Unitary)} & \multicolumn{1}{c}{$G_{AU}$ (Antiunitary) = $\mathcal{T}(G-G_U)$} \\
    \hline
   \{1|0\} : (x, y, z) \to (x, y, z) & \mathcal{T} \{6^+_{001}|0~0~\frac{1}{2}\} : (x, y, z) \to (x-y, x, z+\frac{1}{2}) \\ 
    : (u, v, w) \to (u, v, w) &  : (u, v, w) \to (v, -u-v, -w)\\
   \{3^+_{001}|0\} : (x, y, z) \to (-y, x-y, z) & \mathcal{T} \{6^-_{001}|0~0~\frac{1}{2}\} : (x, y, z) \to (y, -x+y, z+\frac{1}{2}) \\
   : (u, v, w) \to (-u-v, u, w) & : (u, v, w) \to (-u-v, u, -w) \\
   \{3^-_{001}|0\} : (x, y, z) \to (-x+y, -x, z) & \mathcal{T} \{2_{001}|0~0~\frac{1}{2}\} : (x, y, z) \to (-x, -y, z+\frac{1}{2}) \\
   : (u, v, w) \to (v, -u-v, w) & : (u, v, w) \to (u, v, -w) \\
   \{2_{100}|0\} : (x, y, z) \to (x-y, -y, -z) & \mathcal{T} \{2_{210}|0~0~\frac{1}{2}\} : (x, y, z) \to (x, x-y, -z+\frac{1}{2}) \\
   : (u, v, w) \to (u, -u-v, -w) & : (u, v, w) \to (-u-v, v, w) \\
   \{2_{110}|0\} : (x, y, z) \to (y, x, -z) & \mathcal{T} \{2_{120}|0~0~\frac{1}{2}\} : (x, y, z) \to (-x+y, y, -z+\frac{1}{2}) \\
   : (u, v, w) \to (v, u, -w) & : (u, v, w) \to (u, -u-v, w) \\
   \{2_{010}|0\} : (x, y, z) \to (-x, -x+y, -z) & \mathcal{T} \{2_{1\bar{1}0}|0~0~\frac{1}{2}\} : (x, y, z) \to (-y, -x, -z+\frac{1}{2}) \\
   : (u, v, w) \to (-u-v, v, -w) & : (u, v, w) \to (v, u, w) \\
   \{-1|0\} : (x, y, z) \to (-x, -y, -z) & \mathcal{T} \{-6^+_{001}|0~0~\frac{1}{2}\} : (x, y, z) \to (-x+y, -x, -z+\frac{1}{2}) \\
   : (u, v, w) \to (-u, -v, -w) & : (u, v, w) \to (-v, u+v, w) \\
   \{-3^+_{001}|0\} : (x, y, z) \to (y, -x+y, -z) & \mathcal{T} \{-6^-_{001}|0~0~\frac{1}{2}\} : (x, y, z) \to (-y, x-y, -z+\frac{1}{2}) \\
   : (u, v, w) \to (u+v, -u, -w) & : (u, v, w) \to (u+v, -u, w) \\
   \{-3^-_{001}|0\} : (x, y, z) \to (x-y, x, -z) & \mathcal{T} \{m_{001}|0~0~\frac{1}{2}\} : (x, y, z) \to (x, y, -z+\frac{1}{2}) \\
   : (u, v, w) \to (-v, u+v, -w) & : (u, v, w) \to (-u, -v, w) \\
   \{m_{100}|0\} : (x, y, z) \to (-x+y, y, z) & \mathcal{T} \{m_{120}|0~0~\frac{1}{2}\} : (x, y, z) \to (x-y, -y, z+\frac{1}{2}) \\
   : (u, v, w) \to (-u, u+v, w) & : (u, v, w) \to (-u, u+v, -w) \\
   \{m_{110}|0\} : (x, y, z) \to (-y, -x, z) & \mathcal{T} \{m_{210}|0~0~\frac{1}{2}\} : (x, y, z) \to (-x, -x+y, z+\frac{1}{2}) \\
   : (u, v, w) \to (-v, -u, w) & : (u, v, w) \to (u+v, -v, -w) \\
   \{m_{010}|0\} : (x, y, z) \to (x, x-y, z) & \mathcal{T} \{m_{1\bar{1}0}|0~0~\frac{1}{2}\} : (x, y, z) \to (y, x, z+\frac{1}{2}) \\
   : (u, v, w) \to (u+v, -v, w) & : (u, v, w) \to (-v, -u, -w) \\
    \end{tabular}
\end{ruledtabular}
\end{table*}
% ===== =====
\subsection{\label{sec:spinsplitting}Spin splitting without considering SOI}
% ===== Figure: Spin Splitting All Planes =====
\begin{figure*}
	\includegraphics[scale = 0.64]{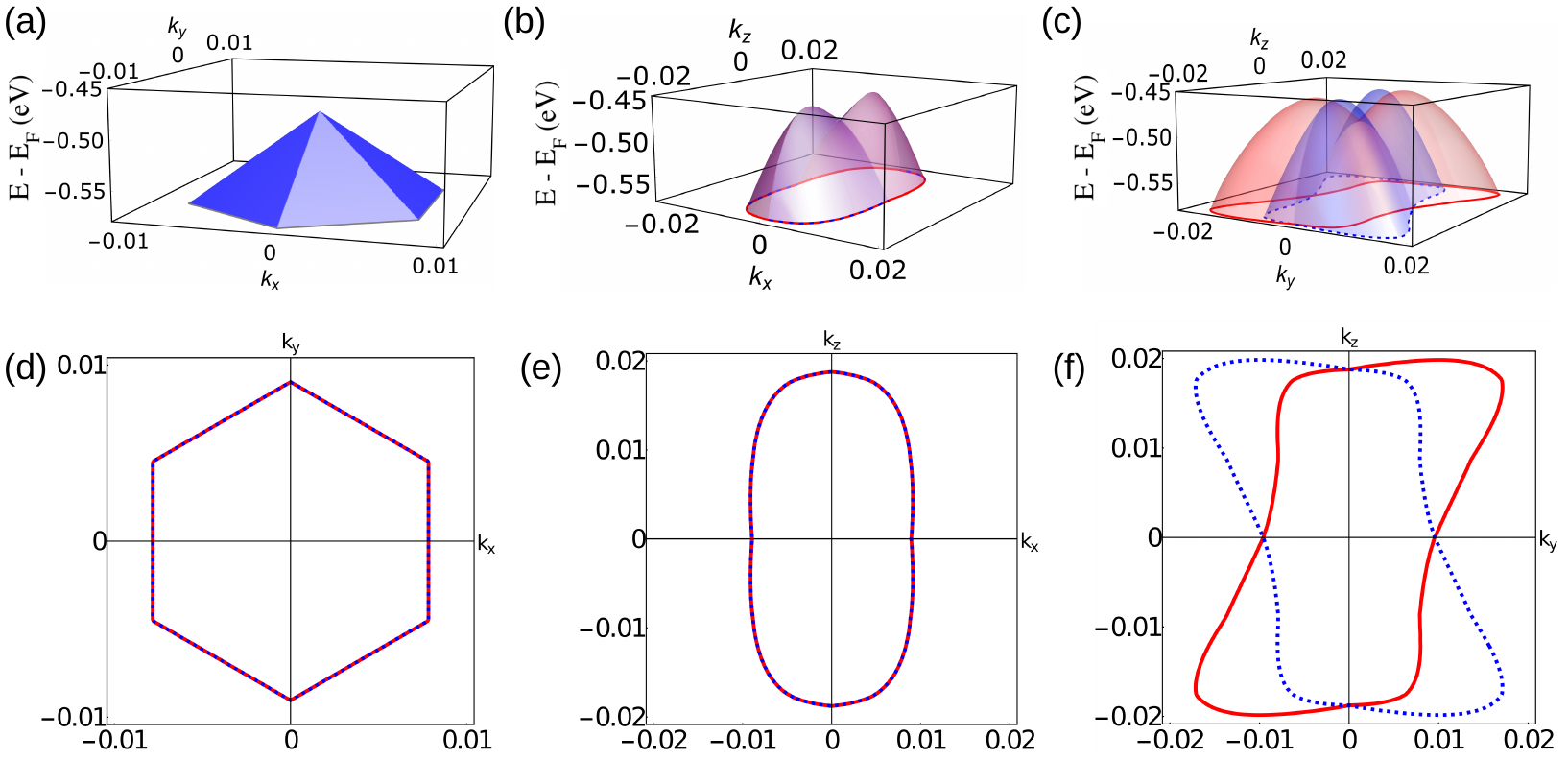}
	\caption{\label{fig:AFMspinSplittingMnTe}The 3D dispersions of the chosen bands in Cartesian coordinate system as functions of ($k_x, k_y$) for $k_z = 0$, ($k_x, k_z$) for $k_y = 0$, and ($k_y, k_z$) for $k_x = 0$ are shown in panels (a), (b), and (c), respectively. Panels (d), (e), and (f) show isoenergetic contours in the Cartesian coordinate system corresponding to the bands at $E - E_F = -0.58$~eV in $k_z = 0$, $k_y = 0$, and $k_x = 0$ planes, respectively.}
\end{figure*}
% ===== =====
Selecting a pair of spin-split bands, as highlighted with thicker lines in \cref{fig:fullpath}, we inspect the 3D band dispersions $\varepsilon(k_x, k_y, 0)$, $\varepsilon(k_x, 0, k_z)$, and $\varepsilon(0, k_y, k_z)$, as shown in \cref{fig:AFMspinSplittingMnTe}(a), \ref{fig:AFMspinSplittingMnTe}(b), and \ref{fig:AFMspinSplittingMnTe}(c), respectively. The corresponding isoenergetic contours at $E - E_F = -0.58$~eV, displayed in \cref{fig:AFMspinSplittingMnTe}(d), \ref{fig:AFMspinSplittingMnTe}(e), and \ref{fig:AFMspinSplittingMnTe}(f) for $k_z = 0$, $k_y = 0$, and $k_x = 0$ planes, respectively. Our results indicate a symmetry-protected spin-degeneracy throughout the $k_z = 0$ and $k_y = 0$ planes, while an enormous spin-splitting in the $k_x = 0$ plane, except at four nodes. The intriguing features of spin-degeneracies and spin-splittings observed here call for a thorough analysis of the magnetic space group symmetries, as presented below.
% ===== Magnetic Space Group =====
\subsection{\label{sec:symmanalysis}Magnetic space group symmetry and spin-splitting}
Considering the magnetic order in the system, time reversal operation becomes relevant for symmetry analysis and can be combined with the group elements to form a magnetic space group (MSG) comprising unitary ($G_U$) and antiunitary ($G_{AU}$) operations \cite{DresselhausGroupTheory08}. Without considering spin-orbit interaction, the MSG of MnTe is $P6'_3/m'mc'$ for the magnetic order shown in \cref{fig:MnTestructure}, with the symmetry operations listed in \cref{tab:MSGSpaceGroup}. Using the appropriate symmetry operations, below we analyze the spin degeneracy and spin splitting of bands in different planes of the Brillouin zone.
% ===== kx-ky plane =====
\subsubsection{$k_x$-$k_y$ planes}
% ===== Figure: kxkyContour =====
\begin{figure*}
	\includegraphics[scale = 0.69]{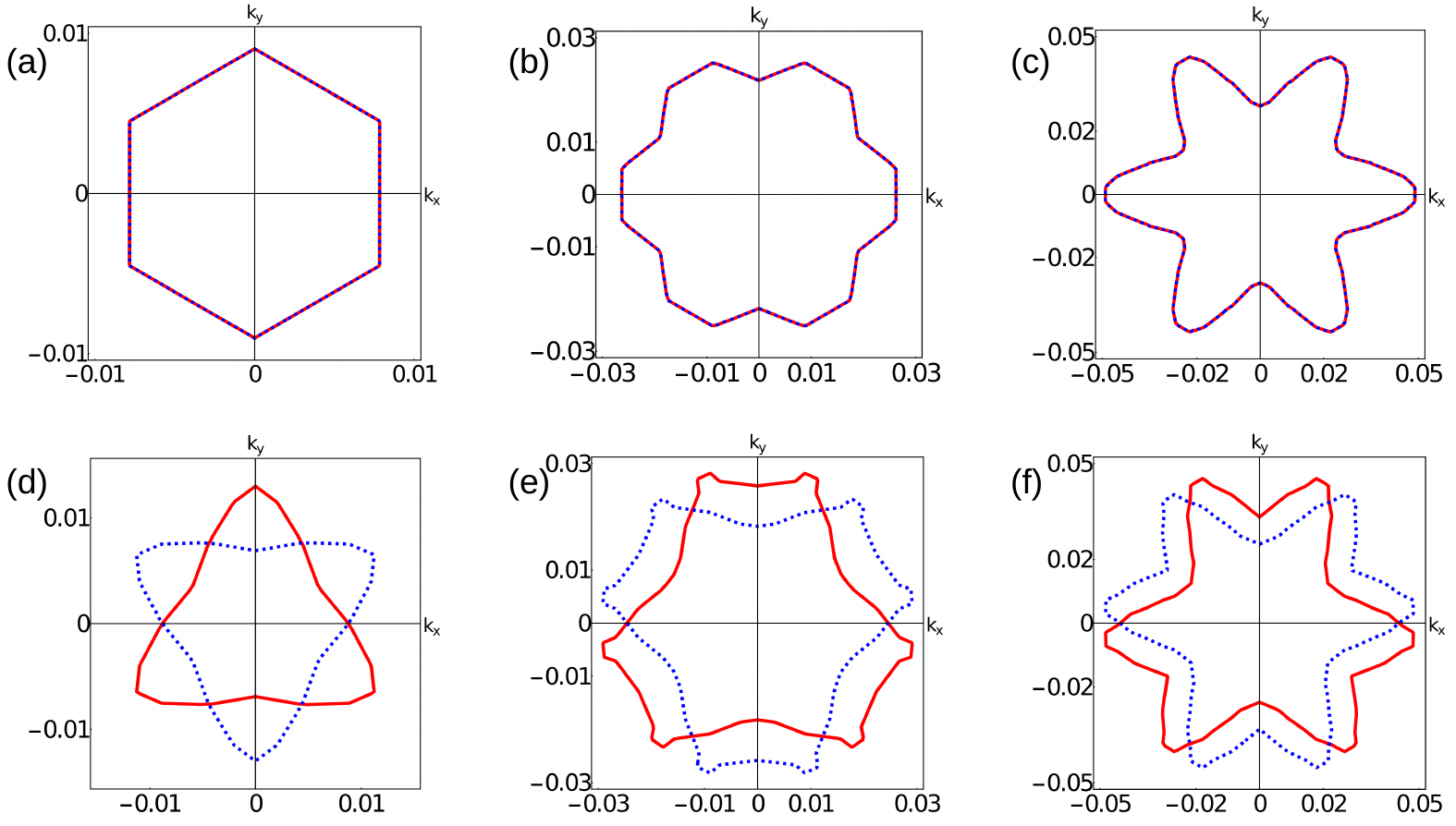}
	\caption{\label{fig:kxky}Panel (a), (b), (c) and (d), (e), (f) shows the isoenergetic contours in Cartesian coordinate system in the $k_z = 0$ and $k_z = 0.0553$~\AA$^{-1}$ planes at $E - E_F = -0.58$~eV, $E - E_F = -0.9$~eV and $E - E_F = -1.1$~eV, respectively.}
\end{figure*}
% ===== =====
The symmetry operations in the magnetic space group (MSG) $P6'_3/m'mc'$ that leave a generic $\vec{k}$-vector ($u, v, 0$) in the $k_z = 0$ plane invariant are (see \cref{tab:MSGSpaceGroup})
\begin{align}
    \{1|0\} &: (u,v,0) \to (u,v,0) \text{ and} \nonumber \\
    \mathcal{T}\left\{2_{001} | 0~0~\frac{1}{2}\right\} &: (u,v,0) \to (u,v,0), \label{eq:kxky}
\end{align}
where the $\vec{k}$-vector is represented in the reciprocal coordinate system. The antiunitary operation $\mathcal{T} \{ 2_{001} | 0~0~\frac{1}{2} \}$ connects the opposite magnetic sublattices, thus ensuring spin-degeneracy throughout the entire $k_z = 0$ plane, as evident from the isoenergetic contours obtained from our DFT results, shown in \cref{fig:kxky}(a), \ref{fig:kxky}(b), and \ref{fig:kxky}(c).

However, considering $k_z$ to be a nonzero constant, the generic $\vec{k}$ vector ($u, v, w = \text{constant}$) does not remain invariant under any of the antiunitary MSG operations in \cref{tab:MSGSpaceGroup}, suggesting the absence of a spin-degeneracy for the $k_x$-$k_y$ planes with a nonzero $k_z$ value. Our DFT results for $k_z = 0.0553$~\AA$^{-1}$ confirm the same, since the opposite-spin isoenergetic contours shown in \cref{fig:kxky}(d), \ref{fig:kxky}(e), and \ref{fig:kxky}(f) do not overlap, except at a few nodes in the Brillouin zone. The $k$-points having symmetry-enforced band degeneracies may be identified by examining how a generic $k$-point ($u, v, w$) transform under the antiunitary symmetry operations that leave $w$ unchanged in $P6'_3/m'mc'$ MSG (see \cref{tab:MSGSpaceGroup}), as illustrated below.
\begin{align}
    \mathcal{T} \left\{2_{1\bar{1}0}|0~0~\frac{1}{2}\right\} &: (u, v, w) \to (v, u, w), \label{eq:finitekz1} \\
    \mathcal{T} \left\{2_{120}|0~0~\frac{1}{2}\right\} &: (u, v, w) \to (u, -u-v, w), \label{eq:finitekz2} \\
    \mathcal{T} \left\{2_{210}|0~0~\frac{1}{2}\right\} &: (u, v, w) \to (-u-v, v, w), \label{eq:finitekz3} \\
    \mathcal{T} \left\{m_{001}|0~0~\frac{1}{2}\right\} &: (u, v, w) \to (-u, -v, w), \label{eq:finitekz4} \\
    \mathcal{T} \left\{-6^-_{001}|0~0~\frac{1}{2}\right\} &: (u, v, w) \to (u+v, -u, w), \text{ and} \label{eq:finitekz5} \\
    \mathcal{T} \left\{-6^+_{001}|0~0~\frac{1}{2}\right\} &: (u, v, w) \to (-v, u+v, w).\label{eq:finitekz6}
\end{align}
The above transformations lead to the following nontrivial conditions for leaving a $\vec{k}$-vector $(u, v, w)$ invariant:
\begin{align}
    u &= v~~\text{from \cref{eq:finitekz1}}, \nonumber\\
    u &= -2v~~\text{from \cref{eq:finitekz2}}, \nonumber\\
    v &= -2u~~\text{from \cref{eq:finitekz3}}. \label{eq:finitekz7}
\end{align}
\cref{eq:finitekz7} suggests symmetry-enforced spin degeneracy at six points in the $k_x$-$k_y$ plane, in perfect agreement with our DFT results shown in \cref{fig:kxky}(d), \ref{fig:kxky}(e), and \ref{fig:kxky}(f) displaying isoenergetic contours for $E - E_F = -0.58$~eV, $-0.9$~eV, and $-1.1$~eV, respectively.
% ===== =====
% ===== ky-kz plane =====
\subsubsection{$k_y$-$k_z$ planes}
% ===== Figure: kykzContour =====
\begin{figure*}
	\includegraphics[scale = 0.69]{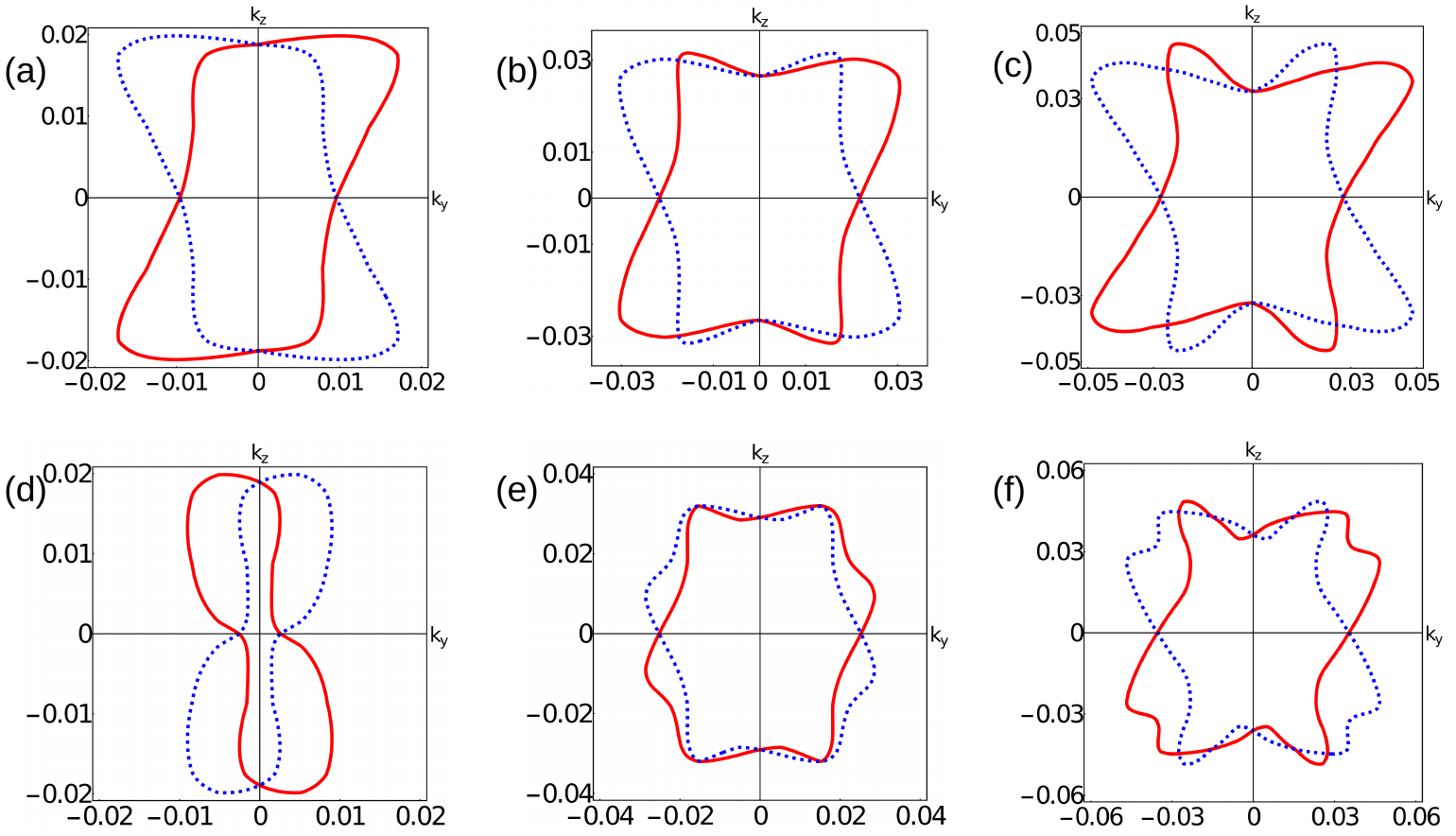}
	\caption{\label{fig:kykz}Panel (a), (b), (c) and (d), (e), (f) shows the isoenergetic contours in Cartesian coordinate system in the $k_x = 0$ and $k_x=0.0559$~\AA$^{-1}$ planes at $E - E_F = -0.58$~eV, $E - E_F = -0.9$~eV and $E - E_F = -1.1$~eV, respectively.}
\end{figure*}
% ===== =====
% ===== Figure: kxkzContour =====
\begin{figure*}
	\includegraphics[scale = 0.68]{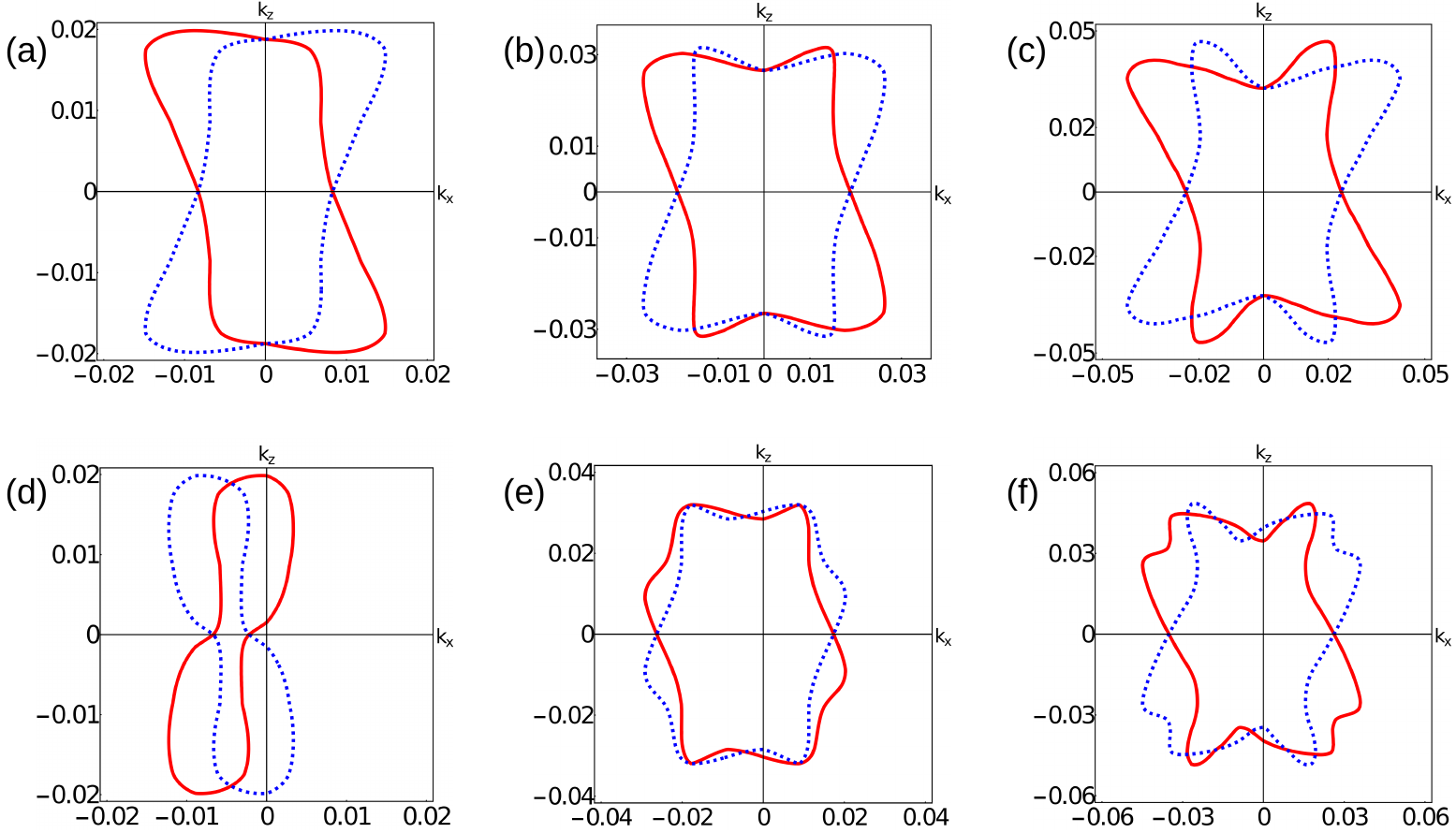}
	\caption{\label{fig:kxkz}Panel (a), (b), (c) and (d), (e), (f) shows the isoenergetic contours in the $v = 0$ and $v = 0.037037$ planes at $E - E_F = -0.58$~eV, $E - E_F = -0.9$~eV and $E - E_F = -1.1$~eV, respectively.}
\end{figure*} 
% ===== =====
Similar to the above, here we analyze the spin splitting in $k_y$-$k_z$ planes. A generic $\vec{k}$-vector $(0, v, w)$ in the $k_x = 0$ plane remains invariant under the following MSG symmetry operations:
\begin{align}
     \{1|0\} &: (0, v, w) \to (0, v, w) \text{ and} \nonumber \\
    \{m_{100}|0\} &: (0, v, w) \to (0, v, w)\label{eq:kykz}
\end{align}
None of these symmetry operations connect the magnetic sublattices with opposite spins, leading to spin splitting. Our DFT results for isoenergetic contours at $E - E_F = -0.58$~eV displayed in \cref{fig:kykz}(a) indeed reveal spin splitting, except for two pairs of nodes at $k_y = 0$ and $k_z = 0$. The isoenergetic contours for $E - E_F = -0.9$~eV and $-1.1$~eV, displayed in \cref{fig:kykz}(b) and \cref{fig:kykz}(c), respectively, indicate more points with spin degeneracy that are not symmetry-protected.

On the other hand, ($u = \text{constant}, v, w$) represents a generic $\vec{k}$-vector for a $k_y$-$k_z$ plane with a nonzero $k_x$ value. The antiunitary symmetry operations in \cref{tab:MSGSpaceGroup} that leave $u$ unchanged upon acting on a $\vec{k}$-vector ($u, v, w$) are
\begin{align}
    \mathcal{T} \left\{2_{001}|0~0~\frac{1}{2}\right\} &: (u, v, w) \to (u, v, -w) \text{ and} \label{eq:finitekx} \\
    \mathcal{T} \left\{2_{120}|0~0~\frac{1}{2}\right\} &: (u, v, w) \to (u, -u-v, w). \label{eq:finitekx1}
\end{align}
Examining \cref{eq:finitekx} we get the condition for $(u = \text{constant},v,w)$ to remain invariant as
\begin{equation}
    w = -w, \text{ requiring } w = 0 = k_z. \label{eq:finitekx2}
\end{equation}
Similarly, \cref{eq:finitekx1} leads to the condition
\begin{equation}
    u + 2v = 0 = k_y ~~(\text{see \cref{eq:condition4}}). \label{eq:finitekx4}
\end{equation}
The isoenergetic contours obtained from our DFT calculations for $k_x = 0.0559$~\AA$^{-1}$ shown in \cref{fig:kykz}(d), \ref{fig:kykz}(e), and \ref{fig:kykz}(f) indicate nodes at $k_z = 0$ and $k_y = 0$ points, in agreement with our analysis.

The additional antiunitary operations acting on ($u = 0, v, w$) point in the $k_x = 0$ plane leaving it in the same plane are (see \cref{tab:MSGSpaceGroup})
\begin{align}
    \mathcal{T} \left\{ m_{001} | 0~0~\frac{1}{2} \right\} &: (0, v, w) \to (0, -v, w), \text{ and} \label{eq:kx01} \\
    \mathcal{T} \left\{ m_{120} | 0~0~\frac{1}{2} \right\} &: (0, v, w) \to (0, v, -w). \label{eq:kx02}
\end{align}
While \cref{eq:kx01} suggests one condition for degeneracy as $v = -v = 0$, implying $k_y = 0$ for $u = 0$ (see \cref{eq:condition4}), \cref{eq:kx02} suggests the other condition as $w = -w = 0$, implying $k_z = 0$ (see \cref{eq:condition5}). Thus, \cref{eq:kx01} and \cref{eq:kx02} enforce no new degeneracy in the $k_x = 0$ plane. Our DFT results, as shown in \cref{fig:kykz}(a), \ref{fig:kykz}(b), and \ref{fig:kykz}(c), are consistent with the symmetry analysis. Besides the degeneracies at $k_y = 0$ and $k_z = 0$, the additional degeneracies found in \cref{fig:kykz}(b), \ref{fig:kykz}(c), \ref{fig:kykz}(e), and \ref{fig:kykz}(f) are not protected by symmetry and may be called accidental degeneracies.
% ===== =====
% ===== kx-kz planes =====
\subsubsection{$k_x$-$k_z$ planes} 
A constant value of $k_y$ represent $k_x$-$k_z$ planes in the Cartesian coordinate system, where \cref{eq:condition4} suggests
\begin{align}
    u + 2v &= \frac{\sqrt{3}a}{2 \pi}k_y = h, \text{ a constant, yielding} \label{eq:finiteky} \\
    u &= h - 2v \text{ for } k_y = \text{constant}. \label{eq:finiteky1}
\end{align}
The antiunitary symmetry operations listed in \cref{tab:MSGSpaceGroup} that leave a generic $\vec{k}$-vector in $k_y = \frac{2 \pi}{\sqrt{3}a}(u + 2v) = \text{constant}$ plane in the same plane (i.e., leave ($u+2v$) invariant) are
\begin{align}
    \mathcal{T} \left\{2_{001}|0~0~\frac{1}{2}\right\} &: (u, v, w) \to (u, v, -w), \text{ and} \label{eq:kyh1} \\
    \mathcal{T} \left\{ m_{120} | 0~0~\frac{1}{2} \right\} &: (u, v, w) \to (-u, u+v, -w). \label{eq:kyh2}
\end{align}
\cref{eq:kyh1} and \cref{eq:kyh2} suggest the following conditions for degeneracy:
\begin{align}
    w &= -w = 0 = k_z, \text{ and} \\
    u &= -u = 0 = k_x.
\end{align}
Thus, we obtain similar degeneracy patterns in $k_y$-$k_z$ and $k_x$-$k_z$ planes.

Considering a special case of $k_y = 0$ plane, implying $u + 2v = 0$, we note that an MSG operation
\begin{equation}
    \mathcal{T} \left\{ 2_{120} | 0~0~\frac{1}{2} \right\} : (u, v, w) \to (u, -u-v, w)
\end{equation}
leaves an arbitrary $\vec{k}$-vector in the plane invariant. Therefore, $k_y = 0$ plane must protect spin-degeneracy, as confirmed from our DFT results shown in \cref{fig:AFMspinSplittingMnTe}(b) and \ref{fig:AFMspinSplittingMnTe}(e).
% ===== =====
% ===== v = constant planes =====
\subsubsection{$v = \text{constant}$ planes}
Following the description of the reciprocal space in \cref{sec:latticevectors}, if we consider a generic $\vec{k}$-point in the $v = \text{constant}$ planes, we identify the following antiunitary symmetry operations from \cref{tab:MSGSpaceGroup} that leave the point in the same plane:
\begin{align}
    \mathcal{T} \left\{ 2_{001} | 0~0~\frac{1}{2} \right\} &: (u, v, w) \to (u, v, -w), \text{ and} \label{eq:finiteky2} \\
    \mathcal{T} \left\{ 2_{210} | 0~0~\frac{1}{2} \right\} &: (u, v, w) \to (-u-v, v, w). \label{eq:finiteky3}
\end{align}
While \cref{eq:finiteky2} leads to $w = -w = 0 = k_z$ as one condition for degeneracy, \cref{eq:finiteky3} indicates the other condition to be $u = -u-v$, implying $u = -v/2$. Thus, we expect band degeneracy at ($-v/2, v, w$) and ($u, v, 0$) points. No other degeneracy condition appears for the special plane $v = 0$; $\mathcal{T} \{ m_{001} | 0~0~\frac{1}{2} \}$ and $\mathcal{T} \{ m_{210} | 0~0~\frac{1}{2} \}$ yield the same degeneracy conditions. \Cref{fig:kxkz}(a), \ref{fig:kxkz}(b), and \ref{fig:kxkz}(c) show isoenergetic contours at different energies for the spin-split bands in $v = 0$ plane, revealing symmetry-enforced degeneracies at $k_x = 0$ and $k_z = 0$ points. On the other hand, \cref{fig:kxkz}(d), \ref{fig:kxkz}(e), and \ref{fig:kxkz}(f) show the isoenergetic contours for the same pair of bands at different energies in $v = 0.037037$ plane, revealing symmetry-enforced degeneracies at $k_x = \frac{2 \pi}{a}u = -\frac{\pi}{a}v$ (see \cref{eq:condition3}) and $k_z = 0$ points. The other degeneracies seen in \cref{fig:kxkz}(b), \ref{fig:kxkz}(c), \ref{fig:kxkz}(e), and \ref{fig:kxkz}(f) are not enforced by any symmetry, therefore, may be called accidental degeneracies.
% ===== =====

We summarize the spin-splitting and spin-degeneracy pattern in different planes in the reciprocal space obtained from our DFT calculations and symmetry analysis in \cref{tab:SplittingPattern}, suggesting spin-splitting throughout the Brillouin zone, except $k_z = 0$ and $k_y = 0$ planes, and the nodal lines identified here.
% ===== Table: Spin Splitting Pattern =====
\begin{table}
    \centering    \caption{\label{tab:SplittingPattern}The spin splitting pattern in different parts of the Brillouin zone of MnTe is summarized here.}
    \begin{ruledtabular}
    \begin{tabular}{lll}
        Plane & Splitting & Nodes \\
        \hline
        $k_z = 0$ & No & $-$ \\
        $k_z = \text{constant} \neq 0$ & Yes & $u = v$, $u = -2v$, $v = -2u$ \\
        & & \\
        $k_y = 0$ & No & $-$ \\
        $k_y = \text{constant} \neq 0$ & Yes & $k_z = 0$, $k_x = 0$ \\
        & & \\
        $k_x = 0$ & Yes & $k_z = 0$, $k_y = 0$ \\
        $k_x = \text{constant} \neq 0$ & Yes & $k_z = 0$, $k_y = 0$ \\
        & & \\
        $v = 0$ & Yes & $k_z = 0$, $k_x = 0$ \\
        $v = \text{constant} \neq 0$ & Yes & $k_z = 0$, $k_x = -\frac{\pi}{a}v$
    \end{tabular}
    \end{ruledtabular}
\end{table}
% ===== =====
% ===== Model Hamiltonian =====
\subsection{\label{sec:Hamiltonian}Effctive two-band model Hamiltonian at the $\Gamma$-point without SOI}
% ===== Table: Hamiltonian =====
\begin{table*}
\caption{\label{tab:Hamiltonian}The transformation properties and the irreducible tensor up to the fourth order are listed according to the generator of the symmetry group of wave vector at $\Gamma$ point for $P6'_3/m'mc'$ MSG.}
\begin{ruledtabular}
    \begin{tabular}{cc.....}
    Symmetrized matrix &  \multicolumn{1}{c}{Irreducible tensor} &  \multicolumn{1}{c}{$\{3^+_{001}|0\}$} & \multicolumn{1}{c}{$\{2_{110}|0\}$}  & \multicolumn{1}{c} {$\{-1|0\}$} & \multicolumn{1}{c}{$\mathcal{T}\{6^+_{001}|0~0~\frac{1}{2}\}$}\\
   \hline
    $\sigma_0$ & $k^2_x+k^2_y, k^2_z$ & 1 & 1 & 1 & 1\\
    $\sigma_z$  & $k_yk_z(3k_x^2-k_y^2)$ & 1 & 1 & 1 & -1\\ 
    \end{tabular}
\end{ruledtabular}
\end{table*}
% ===== =====
After identifying the magnetic symmetry-enforced spin degeneracy in some parts of the Brillouin zone and spin-splitting elsewhere, here we introduce the symmetry-allowed model Hamiltonian to describe the spin-splitting phenomena in the Brillouin zone. We introduce two highly localized spin basis states at the Mn sites according to the antiferromagnetic order. Next, the effective two-band model Hamiltonian at a given $\vec{k}$-point is derived by applying the symmetry constraints imposed by the symmetry group of wave vector on the basis. The symmetry group of wave vector at $\Gamma$ point comprises all the symmetry operations of the magnetic space group $P6'_3/m'mc'$, as listed in \cref{tab:MSGSpaceGroup}. By selectively focusing on the symmetry operations associated with the group's generators, we present the representations and transformation properties of the Pauli vector $\vec{\sigma}$ and the wave vector $\vec{k}$.

The generators of the MSG $P6'_3/m'mc'$ comprise three unitary operations $\{3^+_{001}|0\}$, $\{-1|0\}$, $\{2_{110}|0\}$ and one antiunitary operation $\mathcal{T}\{6^+_{001}|0~0~\frac{1}{2}\}$. The transformation properties of the Pauli matrix and the irreducible tensor operator up to fourth order in $k$ under these generators are listed in \cref{tab:Hamiltonian} that suggests the only invariant term under the symmetry operations as $\sigma_z k_yk_z(3k_x^2-k_y^2)$. The term preserves inversion symmetry, a characteristic of nonrelativistic spin splitting. The Hamiltonian takes the form
\begin{equation}
   H = H_0 + \alpha \sigma_z k_yk_z(3k_x^2-k_y^2), \label{Eq:Hamiltonian}  
\end{equation}
where $H_0$ denotes the Hamiltonian without considering the spin splitting that may be represented by a standard tight-binding Hamiltonian, and $\alpha$ is a constant coefficient. The energy eigenvalues are given as
\begin{equation}
    \varepsilon^{\pm}(\vec{k}) = \varepsilon_0(\vec{k}) \pm \alpha k_yk_z(3k_x^2-k_y^2), \label{eq:eigenvalue}
\end{equation}
where $\varepsilon_0(\vec{k})$ represents the band dispersion without spin splitting. The band dispersion in \cref{eq:eigenvalue} qualitatively explains the spin-splitting features obtained from our DFT calculations, shown in the \cref{fig:kxky}, \ref{fig:kykz} and \ref{fig:kxkz}, as verified in \cref{sec:Eigenvalues}.
% ===== =====
% ===== SOI =====
% ===== Table: Magnetic Space Group w SOI =====
\begin{table}
\caption{\label{tab:MSGSpaceGroupSOI}All symmetry operations of the magnetic space group $Cmcm$, which describes MnTe's magnetic symmetry when spin-orbit interaction is considered, are listed here. The coordinates mentioned here correspond to the primitive lattice vectors as the basis set.}
\begin{ruledtabular}
    \begin{tabular}{:}
    \multicolumn{1}{c}{Symmetry operations} \\
    \hline
   \{1|0\} : (x, y, z) \to (x, y, z) \\
   \{2_{001}|0~0~\frac{1}{2}\} : (x, y, z) \to (-x, -y, z+\frac{1}{2}) \\
   \{2_{110}|0\} : (x, y, z) \to (y, x, -z) \\
   \{2_{1\bar{1}0}|0~0~\frac{1}{2}\} : (x, y, z) \to (-y, -x, -z+\frac{1}{2}) \\
   \{-1|0\} : (x, y, z) \to (-x, -y, -z) \\
   \{m_{001}|0~0~\frac{1}{2}\} : (x, y, z) \to (x, y, -z+\frac{1}{2}) \\
   \{m_{110}|0\} : (x, y, z) \to (-y, -x, z) \\
   \{m_{1\bar{1}0}|0~0~\frac{1}{2}\} : (x, y, z) \to (y, x, z+\frac{1}{2})
    \end{tabular}
\end{ruledtabular}
\end{table}
% ===== =====
\subsection{\label{subsec:MnTeSOI}Altermagnetism upon considering SOI}
Upon incorporating SOI into our DFT calculations, we observe an anisotropy in spin quantization depending on the crystallographic direction, as discussed in Ref.~\cite{RoojAPR23}, with [11$\bar{2}$0] being the preferred direction, leading to $Cmcm$ MSG containing only unitary symmetry operations, as listed in \cref{tab:MSGSpaceGroupSOI}. The corresponding magnetic point group $mmm.1$ consists of three orthogonal mirror planes $m_{001}$, $m_{110}$, and $m_{1\bar{1}0}$. Thus, the magnetic symmetry allows only the magnetic moment components $(m_x, m_x, 0)$ in the hexagonal coordinate system \cite{BilbaoMagNdataMnTe, GallegoJAC16}. $m_{1\bar{1}0}$ and $m_{001}$ mirror planes flip the magnetic moment direction and enforce an antiparallel alignment. Our DFT calculations reveal the antiparallel alignment of local magnetic moments along the easy axis $[11\bar{2}0]$ that cancel each other, leaving no weak ferromagnetic component \cite{RoojPRB25}, consistent with the magnetic symmetry analysis. Therefore, we expect a pure altermagnetic phase with the easy axis along $[11\bar{2}0]$ direction in MnTe. However, experiments reveal a weak ferromagnetism and an anomalous Hall effect in MnTe \cite{KluczykPRB24}, contradicting our expectations. \citet{KluczykPRB24} argue that anomalous Hall conductivity may sustain even if Berry curvature integrates to zero, owing to Berry curvature multipoles \cite{ZhangPRB23}. Further, \citet{MazinARXIV24} attribute the weak, barely measurable ferromagnetism in MnTe to third-order effect in spin-orbit interaction. We note the possibility of an alternative spin quantization direction $[1 \bar{1} 0 0]$ with maximal symmetry in MnTe that would imply a magnetic space group $Cm^{\prime}c^{\prime}m$, leading to a weak ferromagnetism along $\vec{a}_3$ and a first-order anomalous Hall effect \cite{BilbaoMagNdataMnTe, ZhangPRB23}. Besides the possibilities mentioned in Refs.~\cite{KluczykPRB24, MazinARXIV24}, a mixture of $Cmcm$ and $Cm^{\prime}c^{\prime}m$ MSGs may lead to a weak ferromagnetism and a first-order anomalous Hall effect.
% ===== =====
% ===== Conclusion =====
\section{\label{sec:conclusion}Conclusion}
After discussing the symmetry conditions necessary for an altermagnetic spin splitting in MnTe, we analyze the magnetic symmetry in bulk hexagonal MnTe within magnetic space group theory and first-principles density functional theory to understand the spin-splitting behavior without spin-orbit interaction. Our DFT results reveal a sizable spin-splitting along the high-symmetry lines $U \to \Delta$ and $\bar{L} \to \Gamma \to L$. For a critical investigation of the spin-splitting and spin-degeneracies in MnTe without spin-orbit interaction, we conduct a detailed analysis of a pair of bands by examining 3D band dispersions and isoenergetic contours obtained from our DFT calculations and compare them with our symmetry analysis within magnetic space group theory. Our investigation unveils spin degeneracy in the $k_z = 0$ and $k_y = 0$ planes, while spin splitting is observed everywhere else in the Brillouin zone, except for a few nodal lines identified here. We present an insightful analytical model Hamiltonian based on the magnetic space group analysis and the theory of invariants that successfully describes the spin-splitting and spin-degeneracy features in the Brillouin zone. The symmetry analysis presented here for spin-splitting and spin-degeneracy of the electron bands within magnetic space group theory should readily apply to chiral and achiral magnons in MnTe. Further, we investigate the possibility of weak ferromagnetism in MnTe incorporating spin-orbit interaction. Our detailed analysis using magnetic space group symmetry suggests no weak ferromagnetism, which agrees with our DFT results. We also address the experimental report of weak ferromagnetism and anomalous Hall effect in MnTe by discussing different possible scenarios \cite{KluczykPRB24}. Our thorough investigation provides a comprehensive understanding of the altermagnetic spin-splitting behavior in MnTe, offering deep insights into the interplay between magnetic symmetry and the corresponding electronic structure, helping advance electron- and magnon-based spintronics.
% ===== =====
% ========== ACKNOWLEDGMENT ==========
\begin{acknowledgements}
S.R.\ acknowledges a research fellowship from CSIR, India through grant number 09/1020(0157)/2019-EMR-I, India. N.G.\ acknowledges the research fund from SERB, India, through grant number CRG/2021/005320. The use of high-performance computing facilities at IISER Bhopal and PARAM Seva within the framework of the National Supercomputing Mission, India, is gratefully acknowledged.
\end{acknowledgements}
% ===== Appendix: Direct and reciprocal lattice vectors =====
\appendix
\section{\label{sec:latticevectors}Direct and reciprocal lattice vectors}
% ===== Figure: Brillouin zone =====
\begin{figure}
    \includegraphics[scale = 0.2]{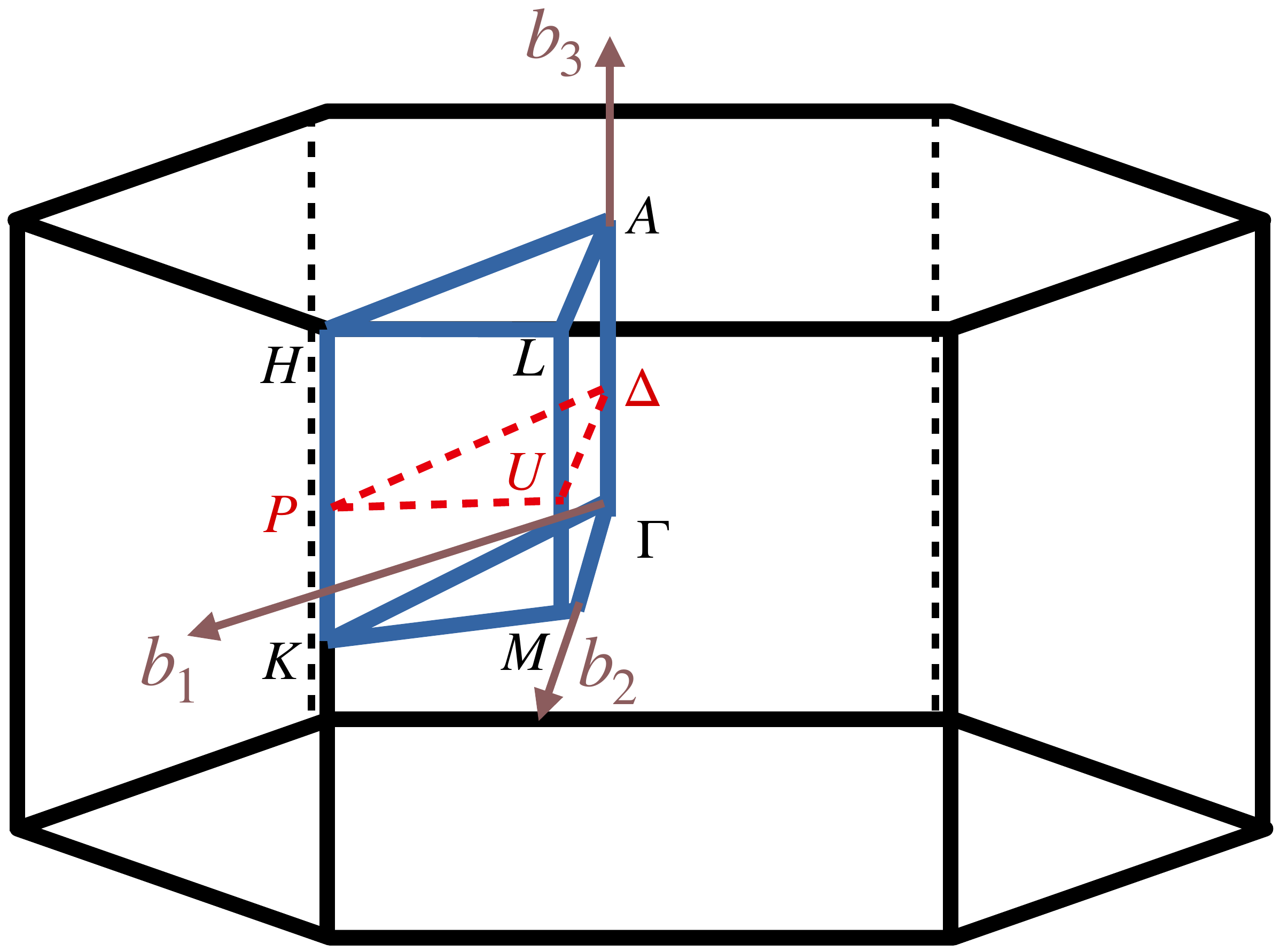}
    \caption{\label{fig:MnTeHBZ}This figure illustrates the hexagonal Brillouin zone with reciprocal lattice vectors, showing all the high symmetry points and lines.}
\end{figure}
Since we address a hexagonal crystal system with non-orthogonal lattice vectors in this article and often need to convert between the reciprocal and Cartesian coordinate systems, here we formally deduce the conversion relations for the particular settings we used for the DFT calculations. The primitive lattice vectors we considered for the hexagonal MnTe structure are
\begin{align}
    \vec{a}_1&=a ~\hat{x}, \nonumber \\
    \vec{a}_2&=-\frac{1}{2}a ~\hat{x}+\frac{\sqrt{3}}{2}a ~\hat{y}, \text{ and} \nonumber \\
    \vec{a}_3&=c ~\hat{z}, \label{eq:primitive}
\end{align}
with the notations having their usual meaning. The corresponding reciprocal lattice vectors, deduced using the relation
\begin{equation}
    \vec{b}_i = 2 \pi \frac{\vec{a}_j \times \vec{a}_k}{\vec{a}_i \cdot (\vec{a}_j \times \vec{a}_k)},
\end{equation}
where $\{i, j, k\} \in \{1, 2, 3\}$ in cyclic order, become
\begin{align}
    \vec{b}_1 &= \frac{2\pi}{a}\left(\hat{x} + \frac{1}{\sqrt{3}} ~\hat{y} \right), \nonumber \\
    \vec{b}_2 &= \frac{4\pi}{\sqrt{3}a} ~\hat{y}, \text{ and} \nonumber \\
    \vec{b}_3 &= \frac{2\pi}{c} ~\hat{z}. \label{eq:reciprocal}
\end{align}
The real and reciprocal lattice vectors have been illustrated in \cref{fig:MnTestructure} and \cref{fig:MnTeHBZ}, respectively. An arbitrary momentum vector denoted by $(u, v, w)$ and $(k_x, k_y, k_z)$ with respect to the reciprocal lattice vectors and in the Cartesian coordinate system, respectively, would imply
\begin{equation}
    u \vec{b}_1 + v \vec{b}_2 + w \vec{b}_3 = k_x~\hat{x} + k_y~\hat{y} + k_z~\hat{z}. \label{eq:condition1}
\end{equation}
Using \cref{eq:reciprocal}, the relation can be written as
\begin{equation}
    \frac{2\pi}{a} u ~\hat{x} + \frac{2\pi}{\sqrt{3} a}(u + 2v) ~\hat{y} + \frac{2 \pi}{c} w ~\hat{z} = k_x~\hat{x} + k_y~\hat{y} + k_z~\hat{z}. \label{eq:condition2}
\end{equation}
Equating the coefficients of $\hat{x}$, $\hat{y}$, and $\hat{z}$ in \cref{eq:condition2}, we obtain
\begin{align}
    k_x &= \frac{2\pi}{a} u, \label{eq:condition3} \\
    k_y &= \frac{2\pi}{\sqrt{3}a}(u + 2v), \text{ and} \label{eq:condition4} \\
    k_z &= \frac{2\pi}{c} w. \label{eq:condition5}
\end{align}
The coordinate transformation may be represented in a matrix form as
\begin{equation}
\begin{pmatrix} k_x \\ k_y \\ k_z \end{pmatrix}
= 2 \pi \begin{pmatrix}
    \frac{1}{a} & 0 & 0 \\
    \frac{1}{\sqrt{3}a} & \frac{2}{\sqrt{3}a} & 0 \\
    0 & 0 & \frac{1}{c}
\end{pmatrix}
\begin{pmatrix}
    u \\ v \\ w
\end{pmatrix}.
\end{equation}
Similarly, the inverse transformation can be represented in a matrix form as
\begin{equation}
\begin{pmatrix} u \\ v \\ w \end{pmatrix}
= \frac{1}{2 \pi} \begin{pmatrix}
    a & 0 & 0 \\
    -\frac{1}{2}a & \frac{\sqrt{3}}{2}a & 0 \\
    0 & 0 & c
\end{pmatrix}
\begin{pmatrix}
    k_x \\ k_y \\ k_z
\end{pmatrix}.
\end{equation}
% ===== =====
% ===== Figure: Model Eigenvalues =====
\begin{figure*}
    \includegraphics[scale = 0.65]{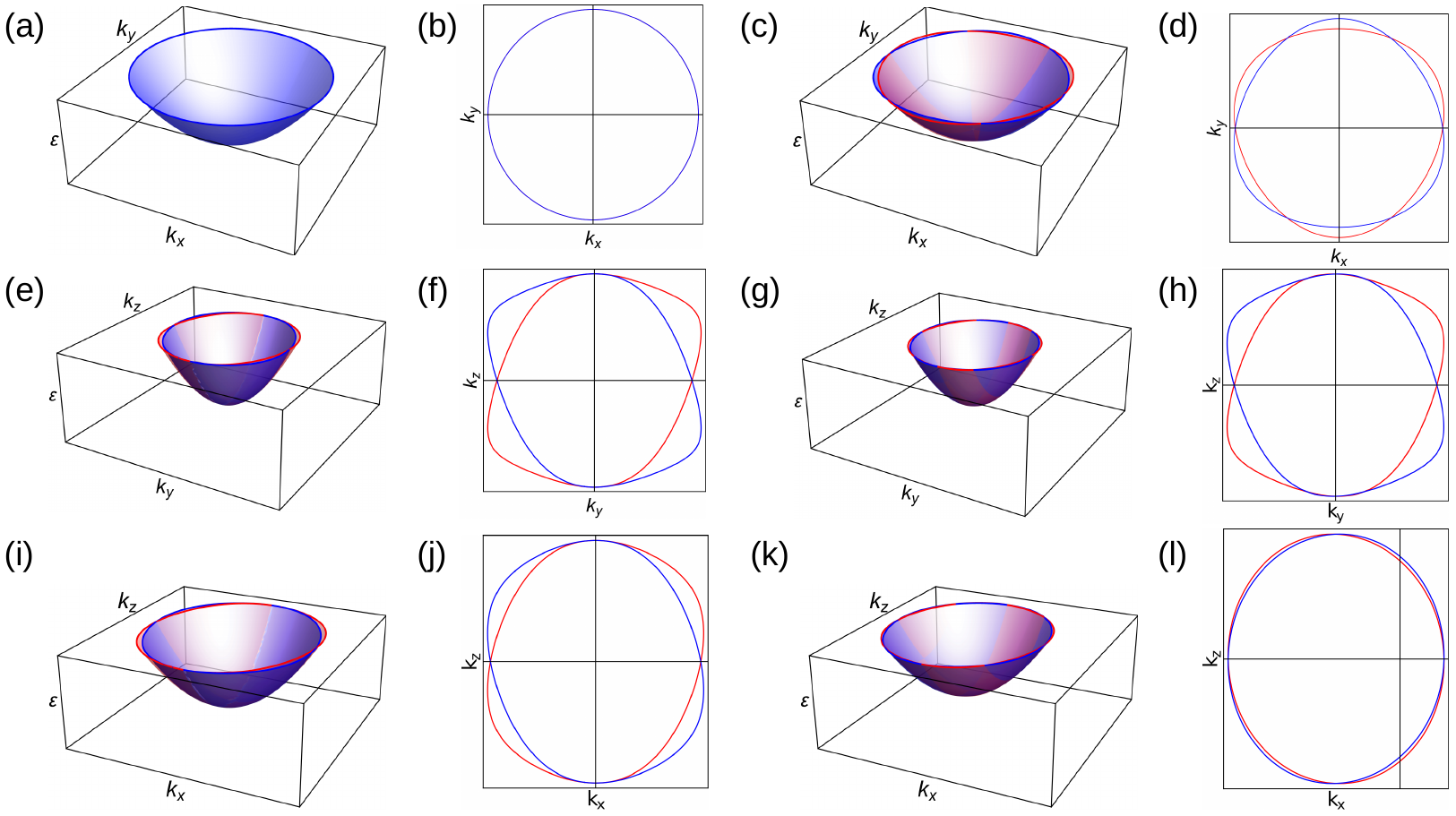}
    \caption{\label{fig:ModelEigenvalues}Panels (a) and (b) show the 3D band dispersion $\varepsilon^{\pm}(k_x, k_y)$ and the corresponding isoenergetic contours, respectively, for the $k_z = 0$ plane, while panels (c) and (d) show similar plots for a $k_z = \text{constant} \neq 0$ plane. Similar plots for the $k_x = 0$ plane and a $k_x = \text{constant} \neq 0$ plane are shown in panels (e), (f) and (g), (h), respectively. Finally, panels (i), (j) and (k), (l) display similar plots for the $v = 0$ plane and a $v = \text{constant} \neq 0$ plane.}
\end{figure*}
% ===== =====
% ===== Symmetry operations in reciprocal space =====
\section{\label{sec:ReciprocalSymmetry}Symmetry operations in the reciprocal space}
Understanding the impact of the symmetry operations in the magnetic space group on an arbitrary momentum vector is key to the symmetry analysis presented in this article. Therefore, we present below how one can find the impact of a symmetry operation in the reciprocal space.

The following relation holds for arbitrary real-space and reciprocal-space lattice vectors $\vec{R}$ and $\vec{G}$, respectively,
\begin{equation}
    \exp{(i\vec{G} \cdot \vec{R})} = 1 = \exp{(2N \pi)}, \label{eq:representation2}
\end{equation}
where $N$ is an integer. A symmetry operation $\alpha$ having real-space matrix representation $U$ acting on a lattice vector $\vec{R}$ transforms it into $\vec{R}^{\prime}$, as represented in tensor notation
\begin{equation}
    R^{\prime j} = U^j_q R^q, \label{eq:representation3}
\end{equation}
assuming a sum over the repeated indices. The same symmetry operation $\alpha$, having a matrix representation $V$ in the reciprocal space, transforms a reciprocal lattice vector $\vec{G}$ as
\begin{equation}
    G^{\prime j} = V^j_p G^p. \label{eq:representation4}
\end{equation}
Since the scalar product of the real and reciprocal lattice vectors must remain invariant under the transformation for \cref{eq:representation2} to be valid, we can write
\begin{align}
    R^{\prime}_j G^{\prime j} &= R_q U^q_j ~ V^j_p G^p = R_q G^q, \text{ implying} \nonumber \\
    U^q_j V^j_p &= \delta^q_p.
\end{align}
Thus, we obtain the desired relation
\begin{equation}
    V = (U^{-1})^T. \label{eq:ConvertOperator}
\end{equation}
% ===== =====
% ===== Model Hamiltonian eigenvalues =====
\section{\label{sec:Eigenvalues}Energy eigenvalues obtained from the model Hamiltonian}
In order to check the validity of the model Hamiltonian derived in \cref{Eq:Hamiltonian}, we plot the corresponding band dispersions and compare them with our DFT results. For simplicity, we take the $H_0$ term in \cref{Eq:Hamiltonian} as the kinetic energy operator that corresponds to a free electron, leading to the energy eigenvalues
\begin{equation}
    \varepsilon^{\pm}(\vec{k}) = \frac{k^2}{2m} \pm \alpha k_yk_z(3k_x^2-k_y^2)
\end{equation}
in the units of $\hbar$. We plot these energy eigenvalues in the forms of 3D band dispersions and isoenergetic contours, as displayed in \cref{fig:ModelEigenvalues}, to compare with our DFT results exhibited in \cref{fig:AFMspinSplittingMnTe}, \ref{fig:kxky}, \ref{fig:kykz}, and \ref{fig:kxkz}, and summarized in \cref{tab:SplittingPattern}. A comparison between our DFT results and the results obtained from this model reveal qualitatively the same pattern for spin-splitting and spin-degeneracies, although the band dispersion looks quite different, owing to our oversimplified assumption for $H_0$ in \cref{Eq:Hamiltonian} being only the kinetic energy operator.
% ===== =====
% ========== BIBLIOGRAPHY ==========
%\bibliography{/Users/nirmal/Documents/bibliography/library,library}

%apsrev4-2.bst 2019-01-14 (MD) hand-edited version of apsrev4-1.bst
%Control: key (0)
%Control: author (8) initials jnrlst
%Control: editor formatted (1) identically to author
%Control: production of article title (0) allowed
%Control: page (0) single
%Control: year (1) truncated
%Control: production of eprint (0) enabled
%
% ====================
\end{document}